\documentstyle[12pt]{article}
\author{Serge Galam}
\date{$\,$\\
Groupe de Physique des Solides\footnotemark[1], Universit\'e Paris 7, Tour 23,\\
2 place
Jussieu,
75251 Paris Cedex 05,
 France\\ (E-mail: galam@gps.jussieu.fr)\\$\,$\\and \\ $\,$\\
Acoustique et Optique de la Mati\`{e}re Condens\'{e}e\footnotemark[2], Universit\'e Paris 6,\\
Tour 13 - Case 86, 4 place Jussieu, 75252 Paris Cedex 05, France\\ $\,$ \\
{\bf (in Press, Physica A, 1996)}}
\addtocounter{footnote}{0}
\footnotetext{Permanent address. Laboratoire associ\'{e} au CNRS (URA n$^{\circ}$ 17)}
\addtocounter{footnote}{1}
\footnotetext{Laboratoire associ\'{e} au CNRS (URA n$^{\circ}$ 800) }
\title{Fragmentation versus Stability \\in\\Bimodal Coalitions }
\begin{document}
\maketitle

\begin{abstract}

Competing bimodal coalitions among a group of actors are discussed.
First, a model from political sciences is revisited.
Most of the model statements are found not to be
contained in the model. Second, a new coalition model is built. It accounts
for local versus
global alignment with respect to the joining of a coalition. The existence
of two competing
world coaltions is found to yield one unique stable distribution of actors.
On the opposite
a unique world leadership allows the emergence of unstable relationships.
In parallel to regular actors which have a clear coalition choice,
``neutral", ``frustrated" and ``risky" actors are produced.
The cold war organisation after world war II is shown to be rather stable.
The emergence of a
fragmentation process from eastern group disappearance is explained as well
as continuing western
group stability.
Some hints are obtained about possible policies to stabilize world
nation relationships. European construction is analyzed with respect to
european stability.
Chinese stability is also discussed.
\end{abstract}
\newpage
\section{Introduction}

Mathematical tools and physical concepts might be a promishing way to
describe social collective
phenomena. Several attempts along these lines have
been made in past years, in particular to study strike process [1],
political organisations [2], group
decision making [3], social impact [4, 5], outbreak of cooperation [6],
power genesis in
groups [7, 8] and stock market [9].

However such an approach should be carefully controlled. A straightforward
mapping
of a physical theory built
for a physical reality onto a social reality could be rather misleading.
It could lead at best to a nice metaphore without
predictability and at worst to a wrong social theory.

Physics has been successful
in describing macroscopic behavior from microscopic properties.
The task here, is to borrow from
physics those techniques and concepts used to
tackle the complexity of aggregations. In parallel the challenge is to build
a collective theory of social behavior along
similar lines, but  within the specific
constraints of the psycho-social reality. The contribution from physics should
thus be restricted to qualitative guidelines for the mathematical modeling
of complex social
realities. Such a limitation does not make the program less ambitious.

Working at the edge of interdisciplinarity between social sciences and
physical sciences
has different inherant dangers for respectively the physicist and the
social scientist.
Coming from the physics side it is to stay in physics using a social
terminology within
a physical formalism. On the opposite, from the other side the danger is to
dress
subjective belief under a pseudo-scientific langage.

In a recent work Axelrod and Bennett used the physical concept of minimum
energy to build a
landscape model (hereafter denoted as AB) of aggregation [10]. Possible
coalitions and choices
various entities can make among them are studied in this Statistical
Physics based model.

In this paper we first position the AB model within the field of
Statistical Physics.
A change in variables and a gauge transformation are shown to map the AB
model exactly
to a $T=0$ finite ferromagnetic Ising system. On this basis most AB
statements are shown to
be misleading with respect to their actual model. They are indeed
confusing a Mattis spin glass and an Edwards-Anderson spin glass [11].

However along above analysis, a new coalition model can indeed be built to
describe
alignment and competition among a group of actors.
The model is found to embody main properties claimed in the AB model.
Temperature-like instabilities are introduced in the model using a concept
of strange actor.

The following of the paper is organized as follows. The second part
contains a review of the
AB model. in Section 3 a new set of variables is shown to map the model
onto a $``zero\,\,
temperature"$ finite Ising ferromagnet. Using this mapping in Section 4
most AB statements are found not compatible with the model. Their model
turns out
to be indeed a Mattis spin glass-like while their comments are drawn from
the physics of an Edwards-Anderson spin glass. Section
5 deals with a presentation of a new model to describe alignment phenomena.
Our model produces
several new features in the dynamics of bimodal coalitions. Nato versus
Warsaw Pact are
analysed. Stability of respectively Europe and China is discussed. Hints to
include instabilities in the
model are presented in Section 6. Last Section contains some concluding remarks.

\section{The Axelrod-Bennett (AB) model}

Axelrod and Bennett (AB ) address the problem of alignment between two
competing coalitions
within a group of $n$ countries [10]. A set of positive variables $\{s_i\}$
accounts for various
actor sizes where index i runs from 1 to n. A pairwise propensity $p_{ij}$
is also considered
among each pairs of actors. It is positive for cases of
cooperation and negative in cases of conflicts. Propensities are assumed to
be symmetric, i.e.,
$p_{ij}=p_{ji}$.

Each actor has then the choice to be in either one of two coalitions. A
distance $d_{ij}$
is then introduced between each pair of actors $i$ and $j$. It is
$d_{ij}=0$ if $i$ and $j$
belong to the same coalition while $d_{ij}=1$ when they belong to different
coalitions.

Given a configuration $X$ of actors, a quantity called ``frustration" (not
to be confused
with the spin glass frustration),
\begin {equation}
F_i=\sum^{n}_{j=1}s_{j}p_{ij}d_{ij}(X)\,,
\end {equation}
is defined for each nation, where the summation is taken over all other
countries including
$i$ itself with $p_{ii}\equiv 0$.
Given a configuration $X$, all country frustrations sum up to
an ``energy",
\begin {equation}
E(X)=\sum_i s_iF_i\,.
\end {equation}
This ``energy" which measures the level of propensity satisfactions is
rewritten,
\begin{equation}
E(X)=\sum_{i>j}^{n}s_{i}s_{j}p_{ij}d_{ij}(X)\,,
\end{equation}
where the sum runs over the $n(n-1)/ 2$ distinct pairs $(i,j)$. Eq.(3) is
the central formula of
AB model.

It is then postulated that actual configuration is the one which minimizes
the energy $E(X)$. There exist by symmetry $2^{n}/ 2$ distinct sets of
alliances since each country has 2 choices for coalition.
Starting from some initial configuration, Axelrod and Bennett treats the
problem numerically.
A dynamics of the system is implemented by single actor coalition flips. An
actor turns to
the competing coalition
only if the flip decreases its local energy. The system has reached its
stable state once
no more flip occurs. Given $\{s_{i}, p_{ij}\}$, the $\{d_{ij}\}$ are thus
obtained minimizing
Eq. (3). Axelrod and Bennett made following statements about their model.
\begin{description}
\item [(a)] Eq. (1) shows ``{\it that the source of conflict with a small
country is
not as important
for determining alignment as an equivalent source of conflict with a large
country}" [10, p. 214].
\item [(b)] The physical concept of frustration [11] is said to be embodied
in their model with ``{\it For example, if there are three nations that
mutually
dislike each other (such
as Israel, Syria and Iraq), then any possible bipolar
 configuration will leave someone frustrated}" [10, p. 217].
\item [(c)] It is stated that alignment can be predicted in real cases with
``{\it Landscape theory begins
with sizes and pairwise
propensities ... to make predictions
about the dynamics of the system}" [10, p. 217].
\end{description}
We will show below (Section 4) that indeed these three statements are
misleading with repect to
AB model content.

\section{The AB model is a $``T=0"$ problem}

The introduction of a new set of variables shows AB model to map onto a
$``T=0"$ finite
Ising ferromagnet. First the two coalitions are denoted respectively by A
and B.
Then a variable $\eta _i$ is associated to each actor.
It is $\eta _i=+1$ if actor $i$ belongs to alliance A
while $\eta _i=-1$ in case it is part of alliance B. From symmetry all
A-members can
turn to coalition B with a simultaneous flip of all B-members to coalition
A.

Given a pair of actors $(i,j)$ their respective alignment is readily
expressed through
the product $\eta _i\eta _j$. The product is $+1$ when $i$ and $j$ belong
to the same coalition
and $-1$ otherwise. Using variables $\{\eta _i\}$, distance $d_{ij}$ can be
recast exactly
under the form,
\begin{equation}
d_{ij}=\frac{1}{2}(1-\eta _i\eta _j)\,,
\end{equation}
and the configuration energy becomes,
\begin{equation}
E(X)=E_0-\frac{1}{2}\sum_{i>j}^nJ_{ij}\eta _i\eta _j\,,
\end{equation}
where
\begin{equation}
J_{ij}\equiv s_is_jp_{ij}\,,
\end{equation}
with $J_{ii}=0$ and
\begin{equation}
E_0=\frac{1}{2}\sum_{i>j}^nJ_{ij}\,,
\end{equation}
is a constant which dependents on initial propensities and sizes of
involved actors
(countries, firms etc).
However this constant is independent of actual coalition actor distribution.
As such it has no effect over the dynamics of shifting coalitions in the
stable state searching.
Dynamics operates trough the expression,
\begin{equation}
H=-\frac{1}{2}\sum_{i>j}^nJ_{ij}\eta _i\eta _j\,,
\end{equation}
which has to be minimized with respect to $\{\eta _i\}$ given $\{J_{ij}\}$.
Eq. (8) turns out to be the Ising model Hamiltonian with competing
interactions [11]. Cooperation
occurs for $J_{ij}>0$ while $J_{ij}<0$ produces conflict.

Since here the system stable configuration minimizes the
energy, the AB model is indeed at the temperature $``T=0"$.
Otherwise when  $``T\neq 0"$ the
free-energy has to be minimized. In practise for a finite system the theory can
tell which coalitions are possible and how many of them exist. But when
several coalitions
have the same energy, it is not possible to predict which one will be the
actual one.

\section{The undressed AB model}

Above three statements (end of Section 2) by Axelrod and Bennett [10] can
be recast as,
{\it Asymmetric size effect}, {\it Frustration effect},
and {\it Alignment prediction}. Unfortunately these statements are
misleading within
standing AB model. Respective proofs follow.
\subsection{Asymmetric size effect}

Though statement of asymmetric size effect sounds reasonable from Eq. (1)
it is indeed not founded.
Dynamics and stable minima are obtained from minimization of the Eq. (3)
energy. Cost
 in ``energy" for having two countries not aligning
according to their propensity, is multiplicative
of both country sizes. Therefore, in case of a
pair of misaligned countries, respectively large and small, the energy cost
is the same
whatever country breaks proper alignment from associated propensity.

\subsection{Frustration effect}

The frustration statement is misleading with respect to both its physics
counterpart and
its meaning in alliances. The physical concept of frustration as introduced by
Toulouse [11] can be defined precisely with the case of
Israel, Syria and Iraq mentioned by Axelrod and Bennett.

We attach
respectively
the labels 1, 2, 3 to each one of the three countries. In case we have equal
and negative exchange interactions
$J_{12}=J_{13}=J_{23}=-J$ with $J>0$, the associated minimum of the energy
(Eq. (8)) is equal to $-J$. However this value of the minimum is realized
for several possible
and equivalent coalitions. Namely for countries (1, 2, 3) we can have
respectively alignments
(A, B, A), (B, A, A), (A, A, B),
(B, A, B), (A, B, B), and
(B, B, A). First 3 are identical to last 3 by symmetry since here what
matters is which countries
are together within the same coalition. The peculiar property is
that the system never gets stable in just one configuration since it costs no
energy to switch from one onto another. This case is an archetype of
frustration.
It means in particular the existence of several ground states with exactly
the same energy.

Otherwise, for non equal interactions the system has one stable
minimum and no frustration occurs within the physical meaning defined above.
The fact that some interactions are not satisfied does not automatically
imply frustration.
Such a situation prevails in cases studied by Axelrod and Bennett, since
minima exist and are
stable. In other words, the fact that the AB model localizes  well defined
minima is the
proof that no frustration is present in the model. Within the framework of
physical models,
the AB model with
the associated numerical propensities used to determine their actual output
is indeed
a Mattis model, i.e., a random site spin glass without frustration.
Since they eventually found one stable minimum, it means aposteriori that
it is possible to find
a set of site variables which will allow a factorisation of their initial
$p_ij$.
However their discussion and statements are based on a random bond spin glass.

We now make this point more quantitative within the present formalism.
Consider a given site $i$.
Interactions with all others sites can be
represented by a field,
\begin {equation}
h_i=\sum^{n}_{j=1}J_{ij}\eta _j\,
\end {equation}
resulting in an energy contribution
\begin {equation}
E_i=-\eta _ih_i\,,
\end {equation}
to the Hamiltonian $H=\frac{1}{2}\sum^{n}_{i=1}E_i$. Eq. (10) is minimum
for $\eta _i$ and $h_i$ having the same sign. For a given $h_i$ there
exists always
a well defined coalition except for $h_i=0$. In this case site $i$ is
``neutral" since then
both coalitions are identical with respect to its local ``energy" which
stays equal to zero.
A neutral site will flip with probability $\frac{1}{2}$. Such a situation
is absent from AB results.

\subsection{Alignement prediction}

The prediction power of AB model is more subtle. Above formulation using
Eqs. (9, 10)
sheds light on what comes really out of AB model. Indeed the ouptput
reduces to the input.
According to AB, the coupling $\{J_{ij}\}$ are given and only one or two
minima are found for the
energy. To make the argument simple without loosing in generality,
we assume there exists only one minimum.
Once the system reaches its stable equilibrium it gets trapped and the
energy is minimum.
At the minimum the field $h_i$ can be calculated for each site $i$ since
$\{J_{ij}\}$  are known as well as $\{\eta _{i}\}$.

First consider all sites which have the value -1. The existence of a unique
non-degenerate minimum makes associated fields  also negative.
We then take one of these sites, e.g. $k$, and
shift its value
from -1 to +1 by simultaneously changing the sign of all its interactions
$\{J_{kl}\}$ where
$l$ runs from 1 to $n$ ($J_{kk}=0$). This transformation gives,
\begin {equation}
\eta _{k}=+1\,{\rm and}\,h_k>0\,,
\end {equation}
instead of,
\begin {equation}
\eta _{k}=-1\,{\rm and}\,h_k<0\,,
\end {equation}
which means that actor $k$ has shifted from one coalition into the other one.

It is worth to emphazise that such systematic shift of propensities of
actor $k$ has no effect
on the others actors. Taking for instance actor $l$, its unique interaction
with actor $k$
is through $J_{kl}$
which did change sign in the transformation. However as actor $k$ has also
turn to the other
coalition, the associated contribution $J_{kl}\eta _{k}$ to field $h_l$ of
actor $l$ is
unchanged.

The shift process is then repeated for each member of actor k former coalition.
Once all shifts are completed there exits only one unique coalition.
Everyone is cooperating
with all others. The value of the energy minimum is unchanged in the
process.

Above transformation demonstrates the $\{J_{ij}\}$ determine the stable
configuration. It shows in particular that given any site configuration, it
always exists a set of
$\{J_{ij}\}$ which will give that configuration as the unique minimum of
the associated energy.
At this stage, what indeed matters is the calculation of propensities. To get
the right output, i.e., the right alignment is not a result of the model,
but instead, a check of
propensity calculation correctness. Therefore AB output (right alignemnt)
reduces to AB input (propensities) making the statement (c) of Sec. II
misleading.

On this basis we can conclude the input, i.e., propensity calculations and
thus associated
$\{J_{ij}\}$ are the relevant and interesting results of
Axelrod and Bennett.

However thes data have to be handle with caution depending on the specific
cases. Indeed above gauge transformation  shows what matters is the sign of
field
$\{h_{i}\}$ and not a given $J_{ij}$ value. A given set of field signs,
positive and negative,
may be realized through an extremely large
spectrum of $\{J_{ij}\}$.

This very fact opens a way to explore some possible deviations from a
national policy.
For instance given
the state of cooperation and conflict of a group of actors,
it is possible to find out limits in
which local pair propensities can be modified without inducing coalition shift.
Some country can turn from cooperation to conflict or the opposite, without
changing the belonging to a given alliance as long as the associated field
sign is unchanged. It
means that a given country could becomes hostile to some former allies,
still staying in the same
overall coalition. One illustration is given by german recognition of
Croatia against the will of
other european partners like France and England, without putting at stake
its belonging to the
European community. The Falklands war between England and Argentina is
another example since
both countries have strong american partnerships.
\section{A new model}

The idea to build an ``energy"-like approach to describe alignment
processes within
a group of actors could be indeed rather powerful. However the proposed AB
model was shown
to have many setbacks. Nevertheless at this stage we are in a position to
develop another
Statistical Physics like model to address the problem of bimodal coalition
phenomena.
Following the AB model we start with a group of $n$ actors and two
competing coalitions $A$ ans $B$.
We keep above notations.
\subsection{Setting the model}

From historical, cultural and economic frames there exit bilateral
propensities $p_{i,j}\equiv G_{i,j}$
between any pair
of countries $i$ and $j$ to either cooperation $(G_{i,j}>0)$, conflict
$(G_{i,j}<0)$
or ignorance $(G_{i,j}=0)$. Each propensity $G_{i,j}$ depends solely on the
pair $(i,\:j)$ itself
and is positive, negative or zero.
Here factorisation over $i$ and $j$ is not possible. Indeed we are dealing
with competing given
bonds or links. It is equivalent to random
bond spin glasses as opposed to Mattis random site spin glasses [12].

Propensities $G_{i,j}$ are somehow local since they don't account for any
global organization
or net.
However coalitions have been known to exist since long ago. To include such
a macro-level
of alignment we consider the case of two competing bimodal coalitions $A$
and $B$ like for instance
western and eastern blocks during the so-called cold war.

Each country has either a natural belonging to one of the two world level
coalitions or not.
A variable $\epsilon _i$ is then attached to actor $i$. It is $\epsilon
_i=+1$ if actor
should be in $A$, $\epsilon _i=-1$ for $B$ and $\epsilon _i=0$ for no
coalition belonging.
Natural belonging is induced by cultural, political and historical
interests. Within the
two world level coalition framework the benefit $J_{i,j}$
gained by exchanges between a pair of countries $(i,\:j)$ is
always positive since sharing resources, informations, weapons is
basically profitable.
Nevertheless a pair $(i,\:j)$ propensity to cooperation, conflict or
ignorance is
$p_{i,j}\equiv \epsilon _i \epsilon _j J_{i,j}$ which can be positive,
negative or zero.
Now we do have a Mattis random site spin glasses [12].

Including both local and macro exchanges result in the pair propensity
\begin{equation}
p_{i,j}\equiv G_{i,j} +\epsilon _i \epsilon _j J_{i,j}\:,
\end{equation}
between two countries $i$ and $j$ with always $J_{i,j}>0$.

An additional variable $\beta_i=\pm 1$ is introduced to account for
benefit from economic and military pressure attached to a given alignment.
It is still
$\beta _i=+1$ in favor of $A$, $\beta _i=-1$ for $B$ and $\beta _i=0$ for
no belonging.
The amplitude of this economical and military interest is measured by a local
positive field $b_i$ which also accounts for the country size and importance.
At this stage, sets $\{\epsilon _i\}$ and $\{\beta _i\}$ are independent.

Actual actor choices to cooperate or to conflict result from the given set
of above quantites.
The associated energy is,
\begin{equation}
H=-\frac{1}{2}\sum_{i>j}^n\{G_{i,j} +\epsilon _i \epsilon _j J_{ij}\}\eta
_i\eta _j
-\sum_{i}^n \beta _ib_i\eta _i \,,
\end{equation}
where $\{\eta _i=\pm 1\}$ are Ising variables which discriminate between
the two coalition choice
with $\eta _i=+1$ for $A$ and $\eta _i=-1$ for $B$.

\subsection{Cold war scenario}

The cold war scenario means that the two existing world level coalitions
generate much stonger
couplings than purely bilateral ones, i.e., $|G_{i,j}|<J_{i,j}$
since to belong
to a world level coalition produces
more advantages than purely local unproper relationship.
In others words local propensities were unactivated since overwhelmed
by the two block trend. The overall system was very stable.
We can thus take $G_{i,j}=0$.
Moreover each actor must belong to a coalition, i. e.,
$\epsilon _i\neq 0$ and $\beta _i\neq 0$.
In that situation local
propensities to cooperate or to
conflict between two interacting countries result from
their respective individual macro-level coalition belongings. the cold war
energy is,
\begin{equation}
H_{CW}=-\frac{1}{2}\sum_{i>j}^n\epsilon _i \epsilon _j J_{ij}\eta _i\eta _j
-\sum_{i}^n \beta _ib_i\eta _i \,.
\end{equation}

\subsubsection{Coherent tendencies}

We consider first the coherent tendency case in which cultural and
economical trends go
along the same coalition, i.e., $\beta _i=\epsilon _i$. Then from Eq. (15)
the minimum of
$H_{CW}$ is unique with all country propensities satisfied.
Each country chooses its coalition  according to its natural belonging,
i.e., $\eta _i=\epsilon _i$.
This result
is readily proven via the variable change $\tau \equiv \epsilon _i \eta _i$
which
turns the energy to,
\begin{equation}
H_{CW1}=-\frac{1}{2}\sum_{i>j}^n J_{ij}\tau _i\tau _j
-\sum_{i}^n b_i\tau _i \,,
\end{equation}
where $J_{i,j}>0$ are positive constants. Eq. (16) is a ferromagnetic
Ising  Hamiltonian in positive symmetry breaking fields $b_i$. Indeed it
has one unique minimum with all $\tau _i=+1$.

The remarkable result here is that the existence of two apriori world level
coalitions is identical
to the case of a unique coalition with every actor in it. It shed light on
the stability of the Cold
War situation where each actor satisfies its proper relationship.
Differences and conflicts appear
to be part of an overall cooperation within this scenario.
Both dynamics are exactly the same since what matters is the existence of a
well
defined stable configuration. However there exists a difference which is
not relevant at this
stage of the model since we assumed $G_{i,j}=0$. However in reality
$G_{i,j}\neq 0$ making the existence of two coalitions to produce a lower
``energy" than
a unique coalition since then, more $G_{i,j}$ can be satisfied.

It worth to notice that field terms $b_i\epsilon _i \eta _i$ account
for the difference in energy cost in breaking a pair proper relationship
for respectively
a large and a small country.
Consider for instance two countries $i$ and $j$ with $b_i=2b_j=2b_0$.
Associated pair energy is
\begin{equation}
H_{ij}\equiv -J_{ij}\epsilon _i \eta _i\epsilon _j \eta _j-2b_0\epsilon _i
\eta _i
-b_0\epsilon _j \eta _j\,.
\end{equation}
Conditions $\eta _i=\epsilon _i$ and $\eta _j=\epsilon _j$ give the
minimum energy,
\begin{equation}
H_{ij}^m=-J_{ij}-2b_0-b_0\,.
\end{equation}
>From Eq. (18) it is easily seen that in case $j$ breaks proper alignment
shifting to
$\eta _j=-\epsilon _j$ the cost in energy is $2J_{ij}+2b_0$. In parallel
when $i$ shifts
to $\eta _i=-\epsilon _i$ the cost is higher with $2J_{ij}+4b_0$. Therfore
the cost in energy
is lower for a breaking from proper alignment by the small country
($b_j=b_0$) than by
the large country ($b_j=2b_0$).
In the real world, it is clearly not the
same for instance for the US to be against Argentina than to Argentina to
be against the US.

\subsubsection{Uncoherent tendencies}

We now consider the uncoherent tendency case in which cultural and
economical trends may go
along opposite coalitions, i.e., $\beta _i\neq \epsilon _i$. Using above
variable change
$\tau \equiv \epsilon _i \eta _i$, the Hamiltonian becomes,
\begin{equation}
H_{CW2}=-\frac{1}{2}\sum_{i>j}^n J_{ij}\tau _i\tau _j
-\sum_{i}^n \delta _i b_i\tau _i \,,
\end{equation}
where $\delta _i \equiv \beta _i \epsilon _i$ is given and equal to $\pm1$.
$H_{CW2}$ is formally identical to the ferromagnetic Ising Hamiltonian in
random fields $\pm b_i$.
However, here the fields are not random.

The local field term $\delta _i b_i\tau _i$ modifies the country field
$h_i$ in Eq. (9) to
$h_i+\delta _i b_i$ which now can happen to be zero.
This change is qualitative since now there exists the possibility to have
``neutrality", i.e.,
zero local effective field coupled to the individual choice. Switzerland
attitude during World
war II may result from such a situation.
Moreover countries which have opposite cultural and economical trends may
now follow their
economical interest against their cultural interest or vice versa.
Two qualitatively different situations may occur.
\begin{itemize}
\item Unbalanced economical power: in this case we have $\sum_{i}^n\delta_i
b_i \neq 0$.

The symmetry is now broken in favor of one of the coalition. But still
there exists only one minimum.

\item Balanced economical power: in this case we have $\sum_{i}^n\delta_i
b_i = 0$.

Symmetry is preserved and $H_{CW2}$ is identical to the ferromagnetic Ising
Hamiltonian
in random fields which has one unique minimum.
\end{itemize}

\subsection{Unique world leader scenario}

Now we consider current world situation where the eastern block has
disappeared. However it
is worth to emphazise the western block is still active as before in this
model. Within
our notations,
denoting $A$ the western alignment,
we have still $\epsilon _i=+1$ for countries which had
 $\epsilon _i=+1$. On the opposite, countries which had
$\epsilon _i=-1$ now turned to either $\epsilon _i=+1$ or
to $\epsilon _i=0$.

Therefore above $G_{i,j}=0$ assumption based on inequality
$|G_{i,j}|<|\epsilon _i\epsilon _j|J_{i,j}$ no longer holds for each pair
of countries.
In particular propensity $p_{i,j}$ becomes equal to $G_{i,j}$ in respective
cases where
$\epsilon _i=0$, $\epsilon _j=0$ and $\epsilon _i=\epsilon _j=0$.

A new distribution of actors results from the collapse of one block. On the
one hand $A$
coalition countries still determine their actual choices according to
$J_{i,j}$. On the other hand
former $B$ coaltion countries are now found to determine their choices
according to
competing links $G_{i,j}$ which did not automatically agree with former
$J_{i,j}$.
This subset of countries has turned from a Mattis random site spin glasses
without frustration
into a random bond spin glasses with frustration. In others world the
former $B$ coalition
subset has jumped from one
stable minimum to a highly degenerated unstable landscape with many local
minima.
This property could be related to the fragmentation process where ethnic
minorities and states
shift rapidly allegiances back and forth while they were part of a stable
structure just
few years ago.

While the $B$ coalition world organization has disappeared, the $A$
coalition world organization
did not change and is still active. It makes $|G_{i,j}|<J_{i,j}$ still
valid for $A$ countries
with $\epsilon _i\epsilon _j=+1$.
Associated countries thus maintain a stable relationship and avoid a
fragmentation process. This result supports a posteriori arguments
against the dissolution of Nato once Warsaw Pact was disolved.

Above situation could also shed some light on the european debate. It would mean
european stability is a result in particular of the existence of european
structures
with economical reality. These structures produce associated propensities
$J_{i,j}$
much stronger than local competing propensities $G_{i,j}$ which are still
there.
In other words european stability would indeed result from
$J_{i,j}>|G_{i,j}|$  and not from either all $G_{i,j}>0$ or all $G_{i,j}=0$.
An eventual setback of the european construction ($\epsilon _i\epsilon
_jJ_{i,j}=0$)
would then automatically yield a
fragmentation process with activation of ancestral bilateral oppositions.

In this model, once a unique economical as well as military world level
organisation exists, each country interest becomes
to be part of it. We thus have $\beta _i=+1$ for each actor. There may
be some exception like Cuba staying almost alone in former $B$ alignment,
but this case will not
be considered here.
Associated Hamiltonian for the $\epsilon_i =0$ subset actor is,
\begin{equation}
H_{UL}=-\frac{1}{2}\sum_{i>j}^n G_{ij}\eta _i\eta _j
-\sum_{i}^n b_i\eta _i \,,
\end{equation}
which is formally equivalent to a random bond Hamiltonian in a field. At
this stage
$\eta _i=+1$ means to be part
of $A$ coalition which is an international structure. On the opposite
$\eta _i=-1$ is to be in a non-existing $B$-coalition which really means to
be outside of $A$.

For small field with respect to interaction the system may still exhibit
physical-like frustration
depending on the various $G_{i,j}$. In this case the system has many minima
with the same energy.
Perpetual instabilities thus occur in a desperate search for an impossible
stability.
Actors will flip continuously from one local alliance to the other. The
dynamics
we are refering to is an individual flip each time it decreases the energy.
We also allow
a flip with probabilty $\frac{1}{2}$
when local energy is unchanged.

It is worth to point out that only strong local fields may
lift fragmentation by putting every actor in $A$-coalition. It can be achieved
through economical help like for instance in Ukrainia. Another way is
military $A$ enforcement
like for instance in former Yugoslavia.

Our results point out that current debate over integrating former eastern
countries within Nato
is indeed relevant to oppose current fragmentation processes. Moreover it
indicated that an
integration would suppress actual instabilities
by lifting frustration.

\subsection{The case of China}

China is an extremely huge country built up from several very large states.
These state
typical sizes are of the order or much larger than most other countries in
the world.
It is therefore interesting to analyse China stability within our model
since it represents a case of simultaneous
Cold war scenario and Unique world leader scenario.

There exists $n$ states which are all part of
a unique coalition which is the chinese central state. Then all
$\epsilon_i=+1$ but $\beta_i =\pm 1$
since some states keep economical and military interest in the ``union"
$(\beta_i =+1)$
while capitalistic advanced rich states contribute more than their share to
the ``union"
$(\beta_i =-1)$. Associated Hamiltonian is,
\begin{equation}
H=-\frac{1}{2}\sum_{i>j}^n\{G_{i,j} + J_{ij}\}\eta _i\eta _j
-\sum_{i}^n \beta _ib_i\eta _i \,,
\end{equation}
where $J_{i,j}>0$ and $G_{ij}$ is positive or negative depending on
each pair of state $(i,\:j)$.

At this point China is one unified country which means in particular that
$J_{i,j}>|G_{ij}|$
for all pair of states with negative $G_{ij}$. Therefore $\eta _i=+1$ for
each state.
Moreover it also implies $b_i<q_iJ_{i,j}$
where $q_i$ is the number of
states state $i$ interacts with. Within this model, three possible scenari
can be oulined with respect to China stability.
\begin{enumerate}
\item China unity is preserved.

Rich states will go along their actual economic growth with the central
power turning to a
capitalistic oriented federative like structure. It means turning all
$\epsilon_i$ to $-1$
with then $\eta_i=\epsilon_i$. In parallel additional development of poor
states is required in order
to maintain condition  $J_{i,j}>|G_{ij}|$ where some $G_{ij}$ are negative.

\item Some rich states break unity.

Central power is unchanged with the same political and economical
orientation making heavier
limitations over rich state development. At some point the condition
$b_i>q_iJ_{i,j}$ may be achieved for these states. These very states will
then get a lower
``energy"
breaking down from chinese unity. They will shift to $\eta _i=-1$ in their
alignment with
the rest of China which has $\eta _j=+1$.

\item China unity is lost with a fragmentation phenomenon.

In this case, opposition among various states becomes stronger than the
central organisational
cooperation with now $J_{i,j}<|G_{ij}|$ with some negative $G_{ij}$. The
situation would become
spin glass-like and
China would then undergo a fragmentation process. Former China would become
a highly unstable part of the world.
\end{enumerate}

\section{The risky actor driven dynamics}

In principle actors are expected to follow their proper relationship, i.e.,
to minimize their local ``energy". In other words, actors
follow normal and usual patterns of decision. But it is well known
that in real life these expectations are sometimes violated. Then
new situations are created with reversal of on going
policies.

To account for such situations we introduce the risky actor. It is an actor
who goes against his well defined interest. It
is different from the frustrated actor which does not have a well defined
interest.
Up to now everything was done at  $``T=0"$. However a risky actor chooses
coalition associated
to $\eta _i=-1$, although its local field
$h_i$ is positive. Therefore the existence of risky actors requires a
$T\neq 0$ situation.
The case of Rumania, having its own independent foreign policy, in former
Warsaw Pact may be an
illustration
of risky actor behavior. Greece and Turkey in the Cyprus conflict may be
another example.

Once $T\neq0$, it is not the energy which has to be minimized but the free
energy,
\begin{equation}
F=U-TS\,,
\end{equation}
where U is the internal energy, now different from the Hamiltonian and
equal to its thermal
average and S is the entropy. To minimize the free energy means stability
of a group of
countries matters on respective size of each coalition
but not, which actors are actually in these
coalitions. At a fixed "temperature" we thus can expect simultaneous shift
of alliances from several
countries as long as the size of the coalition is unchanged, without any
modification in the
relative strenghts. Egypt quitting soviet camp in the seventies and
Afghanistan joining it
may illustrate these non-destabilizing shifts.

Within the coalition frame temperature could be viewed as a way to account
for some risky trend.
It is not possible to know which particular actor will take a chance but
it is reasonable
to assume the existence of some number of risky actors. Temperature would
thus be a way to
account for some global level of risk taking.

Along ideas developped elsewhere [7, 8] we can assume that a level
of risky behavior is
profitable for the system as a whole. It produces surprises which induce to
reconsider some
aspect of coalitions themselves. Recent danish refusal to the signing of
Maastricht agreement
on closer european unity
may be viewed as an illustration of a risky actor. The net effect have been
indeed to turn what
seemed a trivial and apathetic administrative agreement into a deep and
passionated debate
among european countries with respect to european construction.

Above discussion shows implementation of $T\neq 0$ within the present
approach of coalition
should be rather fruitful.
More elaboration is left for future work.

Last but not least, it is worth to mention two actual fields of research
which could
prove useful to our approach at $T\neq 0$. First, simulated annealing
(slowly decreasing temperature)
which helps to find better ground states. And the somehow similar recent
``mutation works" [13].
For our dynamics at $T=0$ studies [14] may turn useful.
\section{Conclusion}

The choice of alliances is probably one of the most crucial questions faced
by social sytems such
as individuals, nations, ethnic minorities, firms and so on. The
Statistical Physics based
approach opens up a fruitful
way to tackle this basic problem.However we showed that partial use of
Physics can be rather
misleading.
Buiding up a model requires to stick to what is contained in the equations
used.

In this paper we have attempted to propose some new notions to construct a
model of bimodal
alliances. We have shown why the cold war organisation after world war II
was rather stable.
It was then found how the eastern group disappearance  has induced the
emergence of a
fragmentation process. Some hints were obtained about possible policies to
stabilize world
nation relationships. The importance of european construction was also
pointed out.

We have
outlined what could be a
dynamics articulated by the presence of actors who can be either "risky" or
"frustrated"
or ``neutral".
A "risky" actor acts against his well defined interest while a "frustrated"
actor having no
well defined interest, acts randomly. Associated effects are expected to be
instrumental in the
building up of alliances.

At this stage, even if the suggested dynamics can be illustrated by some
examples or analogies, our model remains rather primitive. However we feel
it opens up some
possible new road to explore and to forecast international policies. A
deeper investigation
based on precise data is required to both check the validity of our model
and to modify it
to make it more realistic.
\subsection*{Acknowledgments}
I would like to thank S. Moscovici for drawing my attention to [10].
I indebted to D. Stauffer for numerous comments and critical discussions on
the manuscript.


\end{document}